\documentclass[final,3p,times,twocolumn]{elsarticle}

\usepackage{lineno,hyperref}
\modulolinenumbers[5]

\journal{Journal of \LaTeX\ Templates}

\usepackage{amsthm}
\usepackage{amsmath}

\begin{document}

\begin{frontmatter}

\title{Optical Blocking Performance of CCDs Developed for the X-ray Astronomy Satellite XRISM}

\author[d]{Hiroyuki Uchida\corref{correspondingauthor}}
\ead{uchida@cr.scphys.kyoto-u.ac.jp}
\author[d]{Takaaki Tanaka}
\author[d]{Yuki Amano}
\author[d]{Hiromichi Okon}
\author[d]{Takeshi G. Tsuru}
\author[e,f]{Hirofumi Noda}
\author[e,f]{Kiyoshi Hayashida}
\author[e,f]{Hironori Matsumoto}
\author[e]{Maho Hanaoka}
\author[e]{Tomokage Yoneyama}
\author[e]{Koki Okazaki}
\author[e]{Kazunori Asakura}
\author[e]{Shotaro Sakuma}
\author[e]{Kengo Hattori}
\author[e]{Ayami Ishikura}
\author[h]{Hiroshi Nakajima}
\author[b]{Mariko Saito}
\author[b,c]{Kumiko K. Nobukawa}
\author[g]{Hiroshi Tomida}
\author[a]{Yoshiaki Kanemaru}
\author[a]{Jin Sato}
\author[a]{Toshiyuki Takaki}
\author[a]{Yuta Terada}
\author[a]{Koji Mori}
\author[h]{Hikari Kashimura}
\author[i]{Takayoshi Kohmura}
\author[i]{Kouichi Hagino}
\author[j]{Hiroshi Murakami}
\author[k]{Shogo B. Kobayashi}
\author[a]{Yusuke Nishioka}
\author[a]{Makoto Yamauchi}
\author[a]{Isamu Hatsukade}
\author[l]{Takashi Sako}
\author[l]{Masayoshi Nobukawa}
\author[m]{Yukino Urabe}
\author[m]{Junko S. Hiraga}
\author[n]{Hideki Uchiyama}
\author[o]{Kazutaka Yamaoka}
\author[p]{Masanobu Ozaki}
\author[p,q]{Tadayasu Dotani}
\author[e]{Hiroshi Tsunemi}
\author[x]{Hisanori Suzuki}
\author[x]{Shin-ichiro Takagi}
\author[x]{Kenichi Sugimoto}
\author[x]{Sho Atsumi}
\author[x]{Fumiya Tanaka}
\cortext[correspondingauthor]{Corresponding author}


\address[d]{Department of Physics, Kyoto University, Kitashirakawa Oiwake-cho, Sakyo, Kyoto, Kyoto 606-8502, Japan}
\address[c]{Department of Physics, Kindai University, 3-4-1 Kowakae, Higashi-Osaka, Osaka 577-8502, Japan}
\address[a]{Faculty of Engineering, University of Miyazaki, 1-1 Gakuen Kibanadai Nishi, Miyazaki 889-2192, Japan}
\address[b]{Department of Physics, Faculty of Science, Nara Women's University, Kitauoyanishi-machi, Nara, Nara 630-8506, Japan}
\address[e]{Department of Earth and Space Science, Osaka University, 1-1 Machikaneyama-cho, Toyonaka, Osaka 560-0043, Japan}
\address[f]{Project Research Center for Fundamental Sciences, Osaka University, 1-1 Machikaneyama-cho, Toyonaka, Osaka 560-0043, Japan}
\address[g]{Japan Aerospace Exploration Agency, Institute of Space and Astronautical Science, 2-1-1, Sengen, Tsukuba, Ibaraki 305-8505, Japan}
\address[h]{College of Science and Engineering, Kanto Gakuin University, 1-50-1 Mutsuurahigashi, Kanazawa-ku, Yokohama, Kanagawa 236-8501, Japan}
\address[i]{Department of Physics, Tokyo University of Science, 2641 Yamazaki, Noda, Chiba 270-8510, Japan}
\address[j]{Faculty of Liberal Arts, Tohoku Gakuin University, 2-1-1 Tenjinzawa, Izumi-ku, Sendai, Miyagi 981-3193, Japan}
\address[k]{Department of Physics, Tokyo University of Science, 1-3, Kagurazaka, Sinjuku-ku, Tokyo 162-0825, Japan}
\address[l]{Faculty of Education, Nara University of Education, Takabatake-cho, Nara, Nara 630-8528, Japan}
\address[m]{Department of Physics, Kwansei Gakuin University, 2-2 Gakuen, Sanda, Hyogo 669-1337, Japan}
\address[n]{Science Education, Faculty of Education, Shizuoka University, 836 Ohya, Suruga-ku, Shizuoka 422-8529, Japan}
\address[o]{Department of Physics, Nagoya University, Furo-cho, Chikusa-ku, Nagoya, Aichi 464-8602, Japan}
\address[p]{Japan Aerospace Exploration Agency, Institute of Space and Astronautical Science, 3-1-1 Yoshino-dai, Chuo-ku, Sagamihara, Kanagawa 252-5210, Japan}
\address[q]{Department of Space and Astronautical Science, School of Physical Sciences, SOKENDAI (The Graduate University for Advanced Studies), 3-1-1 Yoshino-dai, Chuou-Ku, Sagamihara, Kanagawa 252-5210, Japan}
\address[x]{Hamamatsu Photonics K.K., 1126-1, Ichino-cho, Hamamatsu, 435-8558, Japan}

\begin{abstract}
We have been developing P-channel Charge-Coupled Devices (CCDs) for the upcoming X-ray Astronomy Satellite XRISM, planned to be launched in 2021. 
While the basic design of the CCD camera (Soft X-ray Imager: SXI) is almost the same as that of the lost Hitomi (ASTRO-H) observatory, we are planning to reduce the ``light leakages'' that is one of the largest problems recognized in  Hitomi data.
We adopted a double-layer optical blocking layer on the XRISM CCDs and also added an extra aluminum layer on the backside of them.
We develop a newly designed test sample CCD and irradiate it with optical light to evaluate the optical blocking performance. 
As a result, light leakages are effectively reduced compared with that of the Hitomi CCDs. 
We thus conclude that the issue is solved by  the new  design and that the XRISM CCDs  satisfy  the mission requirement for the SXI.
\end{abstract}


\end{frontmatter}


\section{Introduction}
Charge-Coupled Devices (CCDs) have been widely used as imaging spectrometers in the field of X-ray astronomy, since their first use on the Advanced Satellite for Cosmology and Astrophysics (ASCA), the fourth Japanese X-ray observatory launched in 1993.
Because CCDs are sensitive to visible and ultra-violet light as well, it is required to block such out-of-band light that may increase the background noise level.
For this reason, past X-ray satellite missions installed an optical-blocking filter in front of sensors (e.g., XMM-Newton \cite{Turner2001}; Suzaku \cite{Koyama2007}).
The recent Japanese satellite Hitomi launched in 2016 (formally known as ASTRO-H \cite{Takahashi2018}) introduced an optical blocking layer (OBL) directly deposited on CCDs.
Although the OBL makes the camera system simple and future missions such as Athena \cite{Barbera2015} are planning to adopt a similar design, the Hitomi OBL suffered of light leakages \cite{Nakajima2018}.

We have been developing P-channel CCDs for the next Japanese X-ray observatory, X-Ray Imaging and Spectroscopy Mission (XRISM \cite{Tashiro2018}), planned to be launched in the Japanese fiscal year 2021.
The CCD camera system, soft X-ray imager (SXI\footnote{The SXI and the X-ray Mirror Assembly (XMA) are collectively called Xtend.} \cite{Hayashida2018}) aboard XRISM is required to reduce light leakages.
We therefore changed the design of the OBL as described in the following section. 

This paper is structured as follows. 
Section~\ref{sec:light} describes light leakages found in the Hitomi data and design modifications of the SXI aboard XRISM in terms of the optical blocking performance.
In Section~\ref{sec:test}, we explain the performance of a sample CCD, the design of which is the same as that of the flight model (FM) of the SXI. 
The obtained results are discussed in Section~\ref{sec:ana} and the optical blocking performance of FM CCDs is also summarized in Section~\ref{sec:FM}.
The errors quoted in the text and error bars displayed in the figures represent a 1-$\sigma$ confidence level.
\section{Description of XRISM SXI CCDs in Terms of Optical Blocking Performance}\label{sec:light}
While the basic design of the XRISM SXI is almost the same as that of Hitomi, we are planning to improve several critical points related to  light leakages. 
Since a detailed device description of the CCD is presented by (Hayashida et al. \cite{Hayashida2018} and Kanemaru et al. \textit{submitted to NIMA}),  here we focus on its optical blocking performance.

\subsection{Light Leakages Found in Hitomi Data}\label{sec:ll}
\begin{figure}[t]
\centering
\includegraphics[width=0.97\columnwidth]{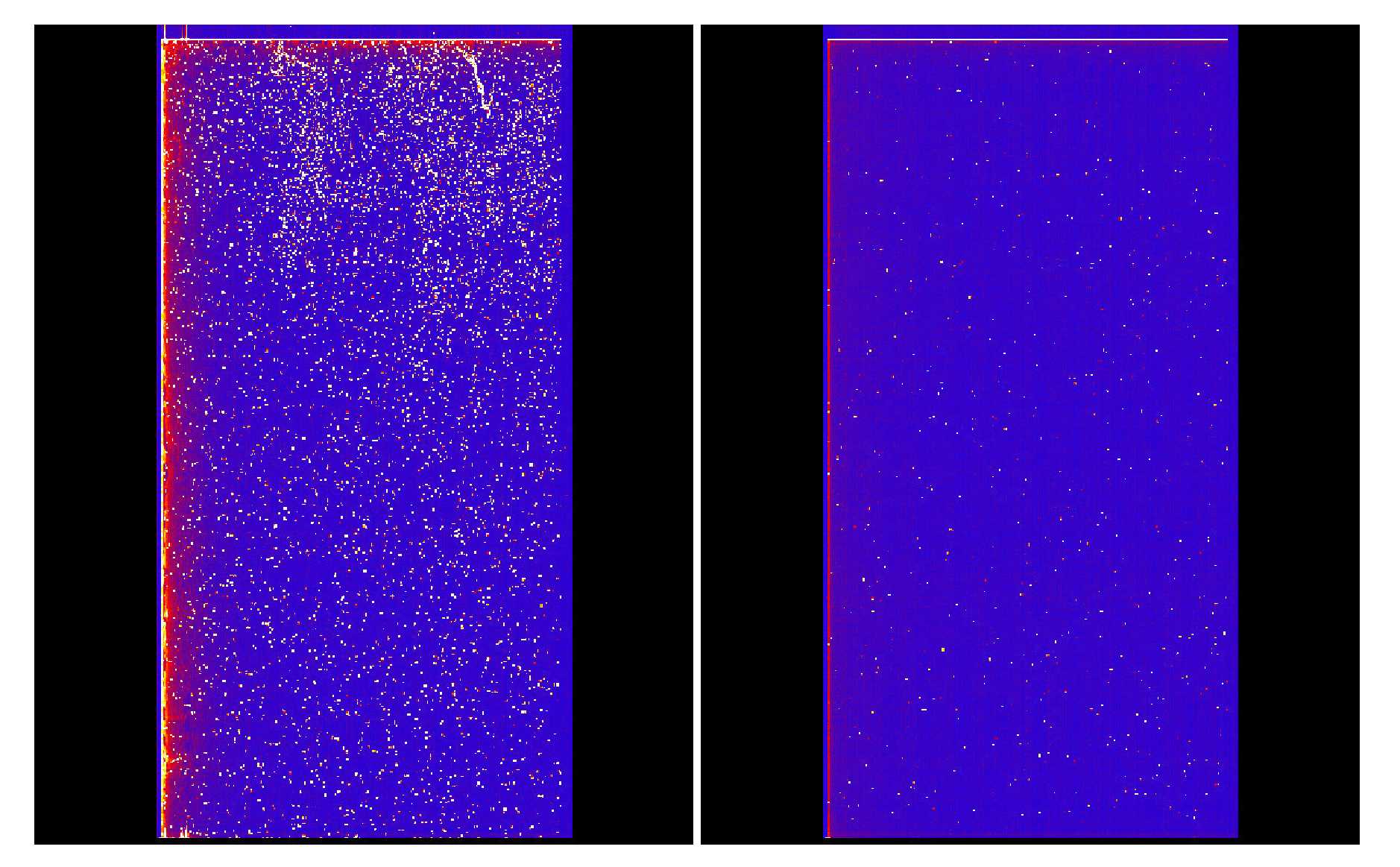}
\caption{LED-irradiated images  with Hitomi (left) and XRISM CCDs (right), which use the same color scale. The bright  region indicates  light leakages. Note that most of the white dots seen in the XRISM CCD image are of X-ray events from a calibration source $^{55}$Fe. The number of pinholes is given in the text.}
\label{fig:image}
\end{figure}

In the case of the CCD instrument aboard Hitomi (similarly named as SXI \cite{Tanaka2018}), an aluminum optical blocking layer (OBL) was deposited on the surface of the CCDs.
The instrument was, however, suffering from significant light leakages.
The phenomenon was observed only during a specific time when the back side of the satellite was toward the day earth (MZDYE \cite{Nakajima2018}).
The main light path is presumed to be holes provided for other instruments in the base panel of the spacecraft, which indicates that the origin of light leakages is the Earth's albedo \textit{i.e.}, reflection of the sunlight from the Earth.
The obtained false signals are categorized into two types according to their event positions: point-like events (pinholes) scattered across the imaging area and end-surface leakage  seen only near the edges of the chips.
The left panel of figure~\ref{fig:image} reproduces the MZDYE light leakages by a ground test with a FM spare CCD for Hitomi.

\subsection{Improvement for XRISM}\label{sec:ll}
While XRISM is designed to close the holes that were the main cause of the problem, we further introduced a fail-safe design so that the CCD instrument itself can minimize the effect of  light leakages.
We took measures ensuring that additional design changes can resolve the above two problems.
Below we summarize the causes of each type of  light leakages and detailed design modifications of the XRISM SXI CCDs.  
Figure~\ref{fig:view} indicates a schematic view of the new design for a XRISM CCD in comparison to that of Hitomi.

\subsubsection{Pinholes}
The origin of ``pinholes'' is voids generated in the aluminum OBL deposited on the Hitomi CCDs with a thickness of 100~nm.
According to a Transmission Electron Microscope image of the OBL, the typical size of these voids is $\sim$1~$\mu$m, which is much smaller than the pixel size of the CCD ($24~\mu\rm{m}\times24~\mu\rm{m}$).
When visible light passing through the voids is scattered in the silicon dioxide layer between the OBL and silicon substrate,  false signals are generated as a single- or multi-pixel event at each position under the pinholes and its surroundings (within $3\times3$ pixels at a maximum).
A similar phenomenon is reported on a ground-base test of CCDs for Origins, Spectral Interpretation, Resource Identification, Security, Regolith Explorer (OSIRIS-REx \cite{Ryu2014, Ryu2017}).
They indicate that it is possibly due to particles or surface irregularities of the silicon substrate under the aluminum layer.

In our case, the main cause of the voids is considered to be a deterioration of the OBL due to a resist stripper used before aluminum vapor deposition. 
A follow-up test revealed that changing the type of the resist stripper can significantly reduce the number of pinholes.
However, even with this change a slight increase of the number of pinholes over year was observed.
We therefore introduced a double-layer aluminum deposition with a thickness of $100+100~{\rm nm}$, so as to decrease the probability of pinhole formation on the OBL (figure~\ref{fig:view}).
Although a reduced quantum efficiency in the soft X-ray band is a trade-off, we confirmed that it still satisfies the requirement of XRISM SXI. 
Note that the double-layer design has also been applied for the light filter of a PNCCD detector aboard extended Roentgen survey with an imaging telescope array (eROSITA \cite{Meidinger2011, Meidinger2012, Granato2012, Granato2013}) launched in July 2019, and no impact on in-orbit data has been reported.

 \subsubsection{End-surface Leakage}
The ``end-surface leakage'' is another problem that we found in the Hitomi data.
Significant light leakages were seen near the field of view of the SXI during MZDYE, which seems similar to that reported by Tsunemi et al. \cite{Tsunemi2010} in day-time data of MAXI/SSC.
They speculate that the light entered through the edge of the MAXI CCD, whereas in our case the boundary around the imaging area of the Hitomi CCD has a dummy pixel region with a width of $250~\mu{\rm m}$, that is  larger than the attenuation length of visible light from the Sun.
We thus conclude that the Hitomi CCD has another light path and that a sidewall coating of a silicon layer (e.g., \cite{Ryu2017}) is less effective for our case.

\begin{figure}[t]
\centering
\includegraphics[width=1\columnwidth]{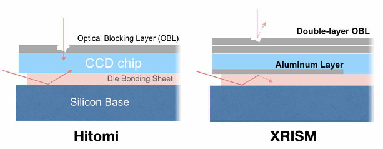}
\caption{Schematic side views of the Hitomi (left) and XRISM (right) CCDs (not to scale). Expected light paths are drawn as the red arrows.}
\label{fig:view}
\end{figure}

As shown in figure~\ref{fig:view}, the lateral structure of the SXI hints at the true origin of the end-surface leakage.
In the design of the Hitomi CCD, a transparent die bonding sheet was used to adhere the silicon substrate and a silicon base. 
It was possible for the light scattered in the sheet to enter the imaging area.
Similar cases are reported by (Bautz et al. \cite{Bautz2016} and Ryu et al. \cite{Ryu2017}) as backside leakage.
To block the entering light, we additionally formed an aluminum filter on the backside electrode (under the passivation) layer of the XRISM CCDs with a width of 2~mm from the edge of the silicon substrate.
The width of the filter is enough longer than the attenuation length of the scattered light, and because its parasitic capacitance is small enough to ignore, it does not affect the charge transfer performance of the CCDs as well.

\section{Performance Test}\label{sec:test}
We evaluate an optical blocking performance of the new CCD with a test sample that has the same design as the FM for XRISM.
The CCD was cooled to $-100~^\circ$C and illuminated with a monochromatic  light using a light emitting diode (LED).
The wavelength was set to $\sim568$~nm, which is near the peak wavelength of the sunlight.
Note that in this experiment we also simultaneously irradiated the CCD with X-rays from an $^{55}$Fe source for the purpose of calibration.

The right panel of Figure~\ref{fig:image} shows a CCD image under the illumination of the above LED obtained with the test sample.
The LED flash duration was 0.1~s for a single frame.
For comparison, we also display an image obtained with a Hitomi FM spare CCD under the same condition in the left panel.
It has a large number of pinholes and also the end-surface leakage with an irregular pattern originating from wrinkles of the die bonding sheet.
These features are the same as those taken with Hitomi, indicating that our ground test reproduces the in-orbit light environment.
The XRISM CCD image shows, on the other hand, a significant reduce both of the number of pinholes and the end-surface leakage region.
The results  demonstrate a substantial improvement of the optical blocking performance.

\begin{figure*}[t]
\centering
\includegraphics[width=0.83\columnwidth]{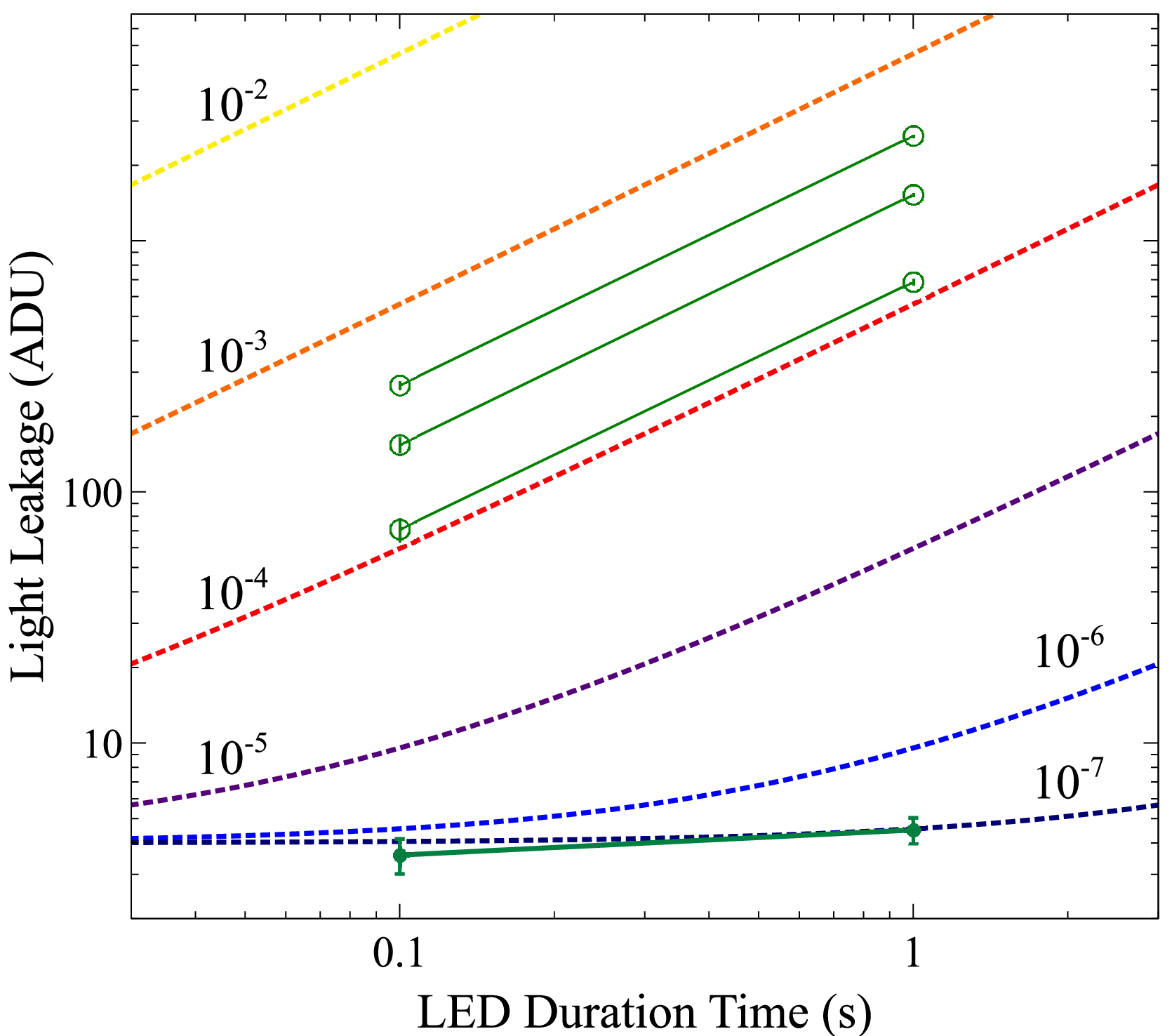}
\includegraphics[width=0.83\columnwidth]{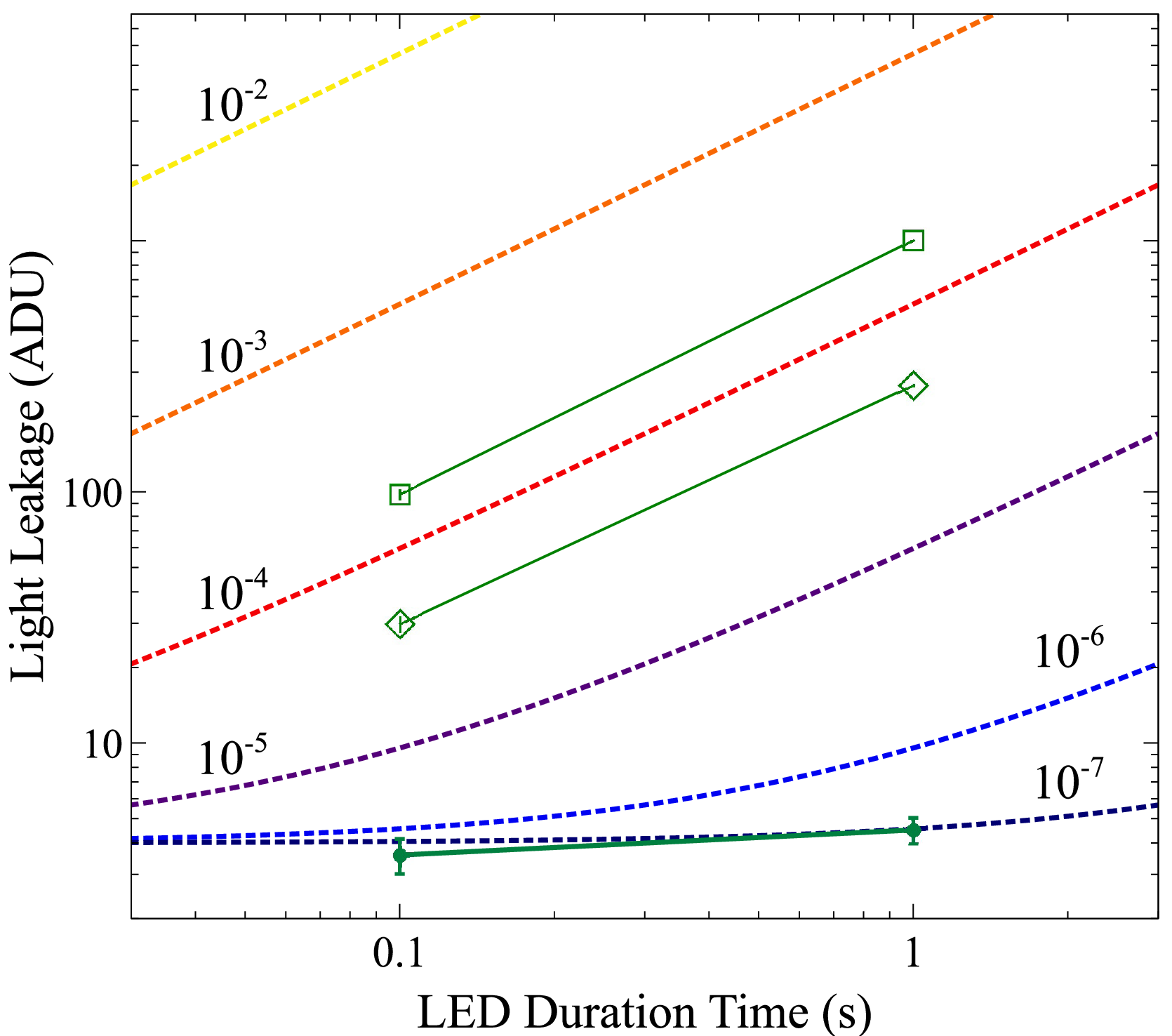}
\caption{Increase of pulse height amplitude caused by light leakages. \textit{Left}: The green open circles correspond to values for pinhole pixels, whereas the filled circles show a result of a normal region. The dotted lines correspond to different transmission factor $T_{\rm X, Y}$ according to equation~(\ref{eq:trans}). \textit{Right}: Same as the left panel, but results of the end-surface leakage are shown. The green squares and diamonds indicate values obtained at the outermost edge ($PH_{\rm 324, 1-200}$) and 3~pixel inward ($PH_{\rm 322, 1-200}$), respectively}
\label{fig:plot}
\end{figure*}

\section{Analysis and Results}\label{sec:ana}
To give a quantitative evaluation of the optical blocking performance of the XRISM CCD, here we define a transmission factor (optical transmission) $T_{\rm X, Y}$ at a pixel position (X, Y) as
\begin{equation}
T_{\rm X, Y} = E_{\rm X, Y}/E_{\rm OBL}, 
\end{equation}
where $E_{\rm X, Y}$ and $E_{\rm OBL}$ are the illuminance of light collected by the pixel and that at the OBL surface.

Given the luminous intensity of the LED $I_{\rm LED}$ is 10~mcd from a specification sheet,  we can estimate the value of $E_{\rm OBL}$ as follows:
\begin{equation}
E_{\rm OBL} \approx 1 \times \left(\frac{I_{\rm LED}}{10~{\rm mcd}}\right) \left(\frac{d}{0.1~{\rm m}}\right)^{-2} \left(\frac{f_{\rm util}}{1}\right)~{\rm lx}, 
\end{equation}
where $d$ and $f_{\rm util}$ are  the distance from the LED to the CCD and the so-called utilization factor, respectively.
While an accurate value of $f_{\rm util}$ depends on the room index and reflectance of the camera chamber, we approximately set $f_{\rm util}\approx1$.
Since the reflectance of inner walls of the aluminum chamber is roughly close to 90\%, our calculation may be an underestimation on $E_{\rm OBL}$.
Also, the LED has a slight directivity, which may give a larger $E_{\rm OBL}$.
These factors result in that the following evaluation of  $T_{\rm X, Y}$ is given under  the most strict condition; smaller $T_{\rm X, Y}$ in reality would be expected than the calculation below.
 
Since 1~lx at $\sim560~{\rm nm}$ corresponds to $1.46\times10^{-3}~{\rm W/m^2}$,  the radiant flux $\Phi_{\rm OBL}$ and the integrated luminous energy $Q_{\rm OBL}$ absorbed per pixel are estimated as
\begin{equation}
\begin{split}
 \Phi_{\rm OBL}  &= 1.46\times10^{-3}  \cdot E_{\rm OBL} \cdot p_{\rm bpx}^2\\
  &=3.36\times10^{-12}~{\rm W}\\
 Q_{\rm OBL}  &= \Phi_{\rm OBL} \cdot t \\
   &= 3.36\times10^{-13}~{\rm J}\\
   &= 2.10\times10^{6}~{\rm eV},
\end{split}
\end{equation}
where $p_{\rm bpx}=48~{\rm \mu m}$ and $t=0.1~{\rm s}$ are the size of a binned pixel and the duration time of the LED illumination, respectively.
Thus, the expected increase of a pulse height amplitude $PH_{\rm X, Y}$ (hereafter, ``light leakages'') in analog-to-digital unit (ADU) can be represented as a function of the photon energy $\varepsilon_\lambda$~(eV) at a wavelength of $\lambda$~(nm) and the CCD gain $g~(e^{-}/{\rm ADU})$:
\begin{equation}
\begin{split}
 PH_{\rm X, Y}  &= PH_0+(T_{\rm X, Y} \cdot Q_{\rm OBL}/\varepsilon_\lambda) \cdot g^{-1}~{\rm ADU}\\
 &= PH_0+(T_{\rm X, Y} \cdot \Phi_{\rm OBL} \cdot t/\varepsilon_\lambda) \cdot g^{-1}~{\rm ADU},
\end{split} 
\end{equation}
where $PH_0$ is an offset due to fluctuations of the dark level between the LED-on and -off data: a few~ADU in our experiment.
The photon energy is fixed at $\varepsilon_{\lambda=568}=2.2~{\rm eV}$ and the gain $g\sim1.67~e^{-}/{\rm ADU}$.

Consequently, if the 0.1-s illumination results in an increase of $PH_{\rm X, Y}$, the transmission factor can be calculated as
\begin{equation}\label{eq:trans}
T_{\rm X, Y} = 1.8\times10^{-6} \cdot (PH_{\rm X, Y}-PH_0) \cdot \left( \frac{t}{0.1~{\rm s}} \right)^{-1}.
\end{equation}
From equation~(\ref{eq:trans}), if an increase of $\sim1000~{\rm ch}$ is caused by a 1-s illumination  at (X, Y), the pixel has a transmission factor $T_{\rm X, Y}\sim1.8\times10^{-4}$.

\begin{figure*}[t]
\centering
\includegraphics[width=0.83\columnwidth]{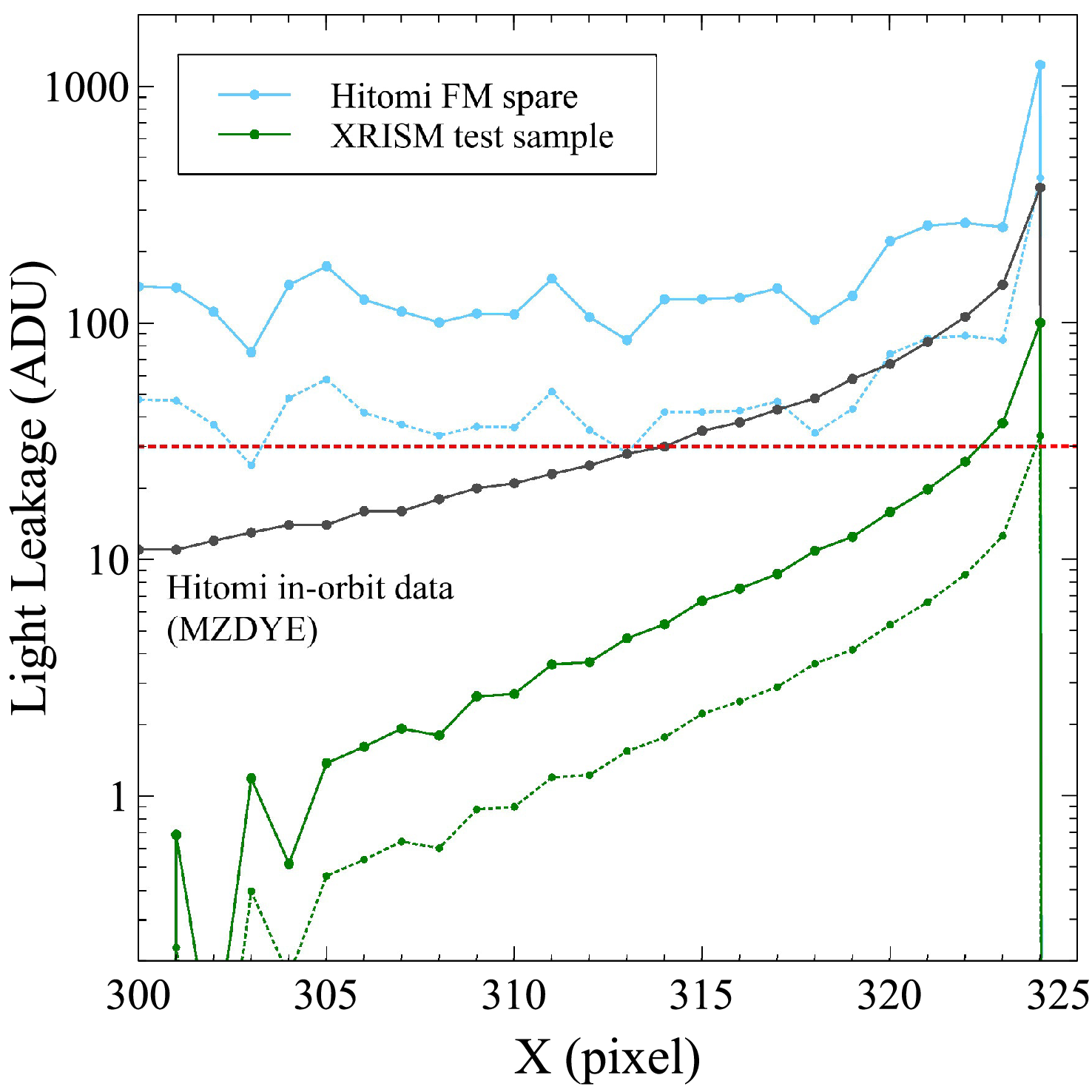}
\includegraphics[width=0.83\columnwidth]{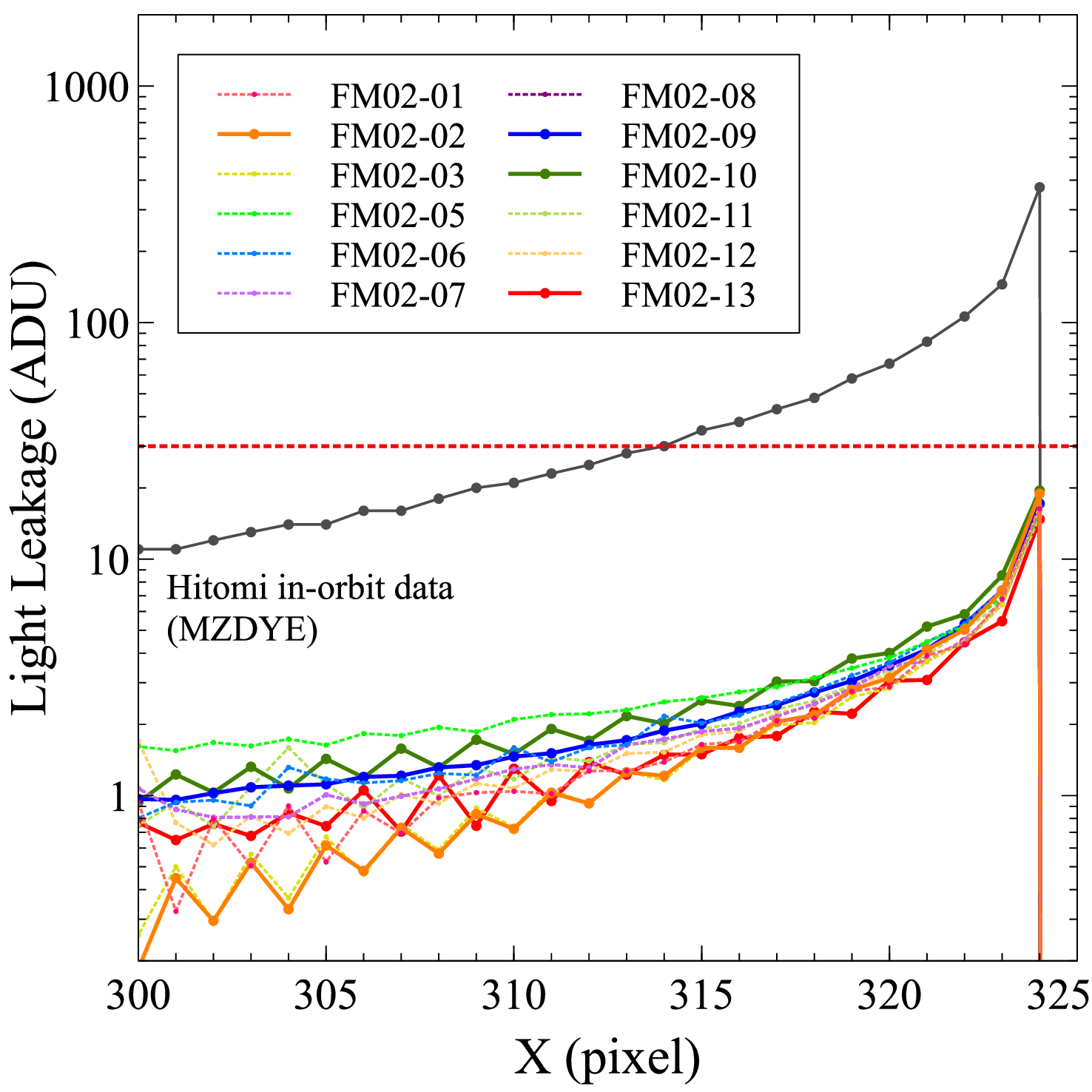}
\caption{\textit{Left}: Pulse height profiles near the edge of the test sample for XRISM (green)  and that of a spare CCD for Hitomi (cyan). The results assuming the in-orbit environment are drawn as the dotted lines with the same colors. The result of Hitomi taken during MZDYE is plotted as the black line.  The mission requirement for the XRISM SXI is drawn as the red dotted line.  \textit{Right}: Same as the left panel but for the XRISM FM CCDs. The results of the selected four and the others are drawn in the solid and dotted lines, respectively.}
\label{fig:projection}
\end{figure*}

\subsection{Pinholes}\label{sec:obppp}
From mission requirements for the XRISM SXI, the number of pinhole pixels should be reduced less than 1.8\% of all.
The definition of the ``pinhole pixels'' given by the XRISM team is those whose transmission factor is $T_{\rm X, Y}<1\times10^{-4}$.
For instance, from equation~(\ref{eq:trans}), an equivalent threshold with the definition is $PH_{\rm X, Y}\sim600~{\rm ch}$ for a 1-s illumination.

To measure  $T_{\rm X, Y}$ for each pixel in the test sample CCD, we irradiated it with the LED light  for various durations (0, 0.1, and 1~s).
Increases of $PH_{\rm X, Y}$ at each pixel were measured by subtracting the average pulse height from LED-on data ($t=0.1, 1~{\rm s}$) to that of LED off ($t=0~{\rm s}$).
In the left panel of figure~\ref{fig:plot}, we plot  obtained  values of light leakages at several example pinhole pixels and that of the surrounding normal region.
The result indicates that the pinhole pixels significantly exceed the threshold transmission  $T_{\rm X, Y}=1\times10^{-4}$, whereas the normal region entirely satisfies the requirement.

The number ratio of the pinhole pixels $n_{\rm pinhole}$ is 0.2\% ($60/25,000~{\rm pixels}$) in the analyzed area.
Since pinhole affected pixels tend to cluster within a few pixels, the real number of pinholes would be less than 0.1\%.
In the same way we calculated  $n_{\rm pinhole}$ of the Hitomi FM CCD as 4.1\% ($1,027/25,000~{\rm pixels}$).
We therefore conclude that  light leakages originated from pinholes were addressed by the double-layer OBL.

\subsection{End-surface Leakage}
It is requested for the XRISM mission to suppress the end-surface leakage up to $T_{\rm X, Y}=1\times10^{-4}$ at 3~pixels inward from the edge.
The right panel of figure~\ref{fig:plot} indicates the average values of light leakages measured at the outermost edge ($PH_{\rm 324, 1-200}$) and that at 3~pixels inward ($PH_{\rm 322, 1-200}$).
Although the illuminance at the end-surface pixel should be somewhat different from $E_{\rm OBL}$, it can still be concluded that almost all inward pixels (${\rm Y}\leq322$) are well below the threshold.

The left panel of figure~\ref{fig:projection} shows comparison between a profile of average $PH_{\rm X, 1-200}$ near the edge of the test sample and that of the Hitomi FM spare CCD.
The end-surface leakage at the outermost column  is measured to be $PH_{\rm 324, 1-200}=100$ and 1231~ADU for XRISM and Hitomi, respectively.
The result indicates a significant improvement on the end-surface leakage roughly by an order of magnitude with the new design.

In the same figure, we also plot the result of Hitomi taken during MZDYE, in which the value of light leakages is $PH_{\rm 324, 1-200}=373$~ADU at the edge.
Given that the Hitomi FM and its spare CCDs have the same optical blocking performance, the result implies that our ground test is in more luminous environment  than Hitomi in orbit, although the configuration (i.e., $f_{\rm util}$) of the camera chamber  is different between the ground test system and the Hitomi SXI. 
If this is the case, the end-surface leakage of the XRISM CCD is almost ignorable even at the outermost edge (the dotted line in figure~\ref{fig:projection}). 

\begin{table}[t]
\caption{Ratio of pinholes in twelve FM CCDs for XRISM.}\label{tab:pinhole}
\begin{center}
\begin{tabular}{lccc}
\hline
\hline
 Numbering & \multicolumn{2}{c}{Ratio of Pinholes}  & SXI Name \\
  \cline{2-3}
& segment AB & segment CD  \\
\hline	   						    		      			  						
 FM02-01   & 0.06\% & 0.04\% & --- \\ 
 FM02-02    & 0.03\% & 0.04\%  & CCD2 \\ 
 FM02-03  & 0.05\% & 0.03\% & ---\\ 
 FM02-05   & 0.27\% & 0.17\% & --- \\ 
 FM02-06   & 0.17\% & 0.30\% & ---\\ 
 FM02-07  & 0.21\% & 0.16\%& ---\\ 
 FM02-08  & 0.21\%&  0.28\%& ---\\ 
 FM02-09  & 0.05\% & 0.03\%  & CCD3 \\ 
 FM02-10   & 0.06\% & 0.04\% & CCD1 \\
 FM02-11   & 0.06\% & 0.04\%& --- \\ 
 FM02-12  & 0.02\% & 0.03\%& ---\\ 
 FM02-13  & 0.02\%&  0.03\% & CCD4\\ 
\hline
\end{tabular}
\end{center}
\end{table}

\section{Optical Blocking Performance of FM CCDs}\label{sec:FM}
Since we  confirmed that the new design  is effective for suppression of both types of  light leakages, we evaluated the optical blocking performance of twelve FM CCDs in a screening test.
In the following analysis, we estimated in-orbit values of light leakages by correcting an expected illuminance based on comparison of the end-surface leakage at the outermost column $PH_{\rm 324, 1-200}$ between the screening test and the Hitomi MZDYE.

The results for the pinholes and the end-surface leakage are summarized in table~\ref{tab:pinhole} and in the right panel of figure~\ref{fig:projection}, respectively.
The ratios of  pinholes for each CCD show that all the CCDs except several segments satisfy the requirement, which will remain unchanged by the end of life of XRISM, even taking into account a possible increase of the pinholes.
As shown in figure~\ref{fig:projection}, we also confirmed that all the pulse height profiles of the FM CCDs are below the given threshold.
The plot indicates that the end-surface leakage as well satisfies the requirement.

We thus conclude that the issues of light leakages found in the Hitomi data are totally solved by changing the optical blocking design and that the XRISM CCDs enough meet  the mission requirement for the SXI.
From these results, we  selected  best four CCDs to be installed in the SXI, also based on several criteria such as the energy resolution and charge transfer efficiency (Yoneyama et al. \textit{submitted to NIMA}).

\section*{Acknowledgements}
This work is supported by JSPS KAKENHI Grand Numbers JP19K03915 (H.U.), JP15H02090 (T.G.T.), JP20H01947 (H.N.), JP16H03983 (K.M.), JP18H01256 (H.N.), and JP17K14289 (M.N.).

\bibliography{HSTD12}


\end{document}